\documentclass[12pt]{iopart}
\usepackage{color}
\usepackage{amssymb}
\usepackage{cite}
\usepackage{graphics}

\begin{document}
\bibliographystyle{iopart-num} 

\title[On Convergence of Born Series]{On the Convergence of the Born
  Series in Optical Tomography with Diffuse Light}

\author{Vadim A. Markel\footnote[1]{E-mail:vmarkel@mail.med.upenn.edu}}

\address{Departments of Radiology and Bioengineering, University of
  Pennsylvania, Philadelphia, PA 19104}

\author{John C. Schotland\footnote[2]{E-mail:schotland@seas.upenn.edu}}

\address{Department of  Bioengineering, University of
  Pennsylvania, Philadelphia, PA 19104}

\begin{abstract}
  We provide a simple sufficient condition for convergence of Born
  series in the forward problem of optical diffusion tomography.  The
  condition does not depend on the shape or spatial extent of the
  inhomogeneity but only on its amplitude.
\end{abstract}

\date{\today} 
\submitto{Inverse Problems}

\maketitle

\section{Introduction}
\label{sec:intro}

Many inverse scattering problems in imaging are known to be nonlinear.
Physically, this is a manifestation of the fact that the probing waves
do not propagate via well-defined trajectories. When such trajectories
do exist, the inverse problem can usually be linearized as is the
case, for example, in single-energy computed x-ray tomography.  If the
probing waves experience scattering and trajectories can not be
defined in principle, nonlinearity of the inverse problem is
practically unavoidable.

Mathematically, the nonlinearity of the inverse problem can be
understood as the nonlinear dependence of the measured signal on the
quantity of interest. In the case of optical tomography (OT), the
measured signal is the intensity of light exiting from a
highly-scattering sample and the the quantities of interest are the
absorption and the scattering coefficients. The nonlinear nature of
the dependence of OT measurements on these coefficients is
well-known~\cite{arridge_99_1,gibson_05_1}.

Practical approaches to solving nonlinear inverse problems can be
divided into two broadly defined classes of iterative and analytic
methods. Iterative methods, including the
Newton-type~\cite{arridge_99_1,roy_01_1,klose_03_1} and
Bayesian~\cite{ye_99_1} methods, seek to optimize a certain cost
function according to an iterative rule which typically requires
solving the forward problem at each iteration step. The advantage of
iterative methods is their generality, since they do not require
knowledge of the analytical structure of the forward operator.
Instead, the forward problem is solved at each iteration step
numerically. Methods of the second class rely on some analytical
manipulations with the forward operator. This includes various
approximate linearization schemes which, generally, work only for weak
inhomogeneities and methods based on functional series expansions.
Thus, image reconstruction algorithms based on an inverse scattering
series were proposed in geophysics (inverse scattering of seismic
waves)~\cite{weglein_97_1,weglein_03_1}, in optical near-field
imaging~\cite{panasyuk_06_2}, and in OT~\cite{markel_03_2}.

While little is presently known about the convergence of the inverse
series, a number of results on convergence of the forward series has
been obtained. In quantum-mechanical scattering theory, Bushell has
shown that the Born series converges if the potential is too shallow
to support at least one bound state~\cite{bushell_72_1}.  Colton and
Kress have studied the convergence of the Born series for the scalar
wave equation in an infinite space~\cite{colton_book_98}. In
particular, as part of the proof of Theorem 8.4 of
Ref.~\cite{colton_book_98}, it is shown that the Born series converges
if the susceptibility $\eta_({\bf r}) = n^2({\bf r}) - 1$ [$n({\bf r})$
being the refractive index] is bounded by $\vert \eta({\bf r}) \vert <
2/(ka)^2$, where $k=\omega/c$ is the wave number and $a$ is the radius
of the smallest sphere that contains the support of $\eta({\bf r})$.
We note that Bushell's convergence condition is indirect and,
therefore, difficult to use.  The convergence condition of Colton and
Kress is applicable to functions of compact support but is not useful
at all in the limit $ka \rightarrow \infty$. In addition, it is only
applicable to scattering by a potential in free space. In this paper,
we show that, in the case of the diffusion equation used in OT, a
simple condition for convergence of the Born series can be obtained
independently of the medium boundaries. A remarkable property of this
condition is that it is also independent of shape or support of the
inhomogeneity but only depends on its amplitude. Thus, we show that
the forward series expansion for the Green's function of the diffusion
equation in powers of absorptive inhomogeneity $\delta\alpha({\bf r})$
(the absorption coefficient is decomposed as $\alpha({\bf r}) =
\alpha_0 + \delta\alpha({\bf r})$ where $\alpha_0$ is a constant)
always converges if

\begin{equation}
\label{1}
\vert \delta\alpha({\bf r}) \vert \leq \alpha_0 \ .
\end{equation}

\noindent
A similar condition is obtained for the diffusion coefficient $D({\bf
  r}) = D_0 + \delta D({\bf r})$. We argue that the independence of
this condition on the shape or spatial extent of the inhomogeneity is
a consequence of the exponential decay of diffuse waves which results
in weak long-range interactions. This argument will be made more
precise in Section~\ref{sec:colton} below and illustrated numerically
in Section~\ref{sec:num}.

The convergence condition (\ref{1}) is obtained independently of
restrictions on the support of the inhomogeneities or of the nature of
medium boundaries. However, if the support of the inhomogeneity is
contained in a finite ball of radius $a$ and the system is embedded in
an infinite homogeneous medium, we can repeat the arguments of
Ref.~\cite{colton_book_98} for the diffusion equation and obtain an
even sharper condition on $\delta\alpha({\bf r})$. Namely, we will
show that, for absorbing inhomogeneities and under the conditions
stated above, the Born series converges if

\begin{equation}
\label{2}
\delta\alpha({\bf r}) < \frac{\alpha_0}{1 - (1 + k_da)\exp(-k_d a)} \
,
\end{equation}

\noindent
where $k_d = \sqrt{\alpha_0/D_0}$ is the diffuse wave number (the
analog of the wave number $k$ of the scalar wave equation). It can be
seen that in the limit $k_d a \rightarrow \infty$, we reproduce the
condition $\delta\alpha < \alpha_0$, while in the limit $k_d a
\rightarrow 0$, we reproduce Colton and Kress' condition $\delta\alpha
< 2\alpha_0/(k_d a)^2$ (note that $\delta\alpha/\alpha_0$ is the
direct analog of the susceptibility $\eta$ of the scalar wave equation
considered by Colton and Kress).

The paper is organized as follows. In Section~\ref{sec:Born} we define
the problem of OT, review the mathematical formalism that leads to the
Born series expansions and introduce the relevant notation. In
Sections~\ref{sec:convergence_alpha} and \ref{sec:convergence_D} we
obtain the convergence condition of the type (\ref{1}) for absorbing
and scattering inhomogeneities, respectively. In Section
\ref{sec:colton} we generalize the Colton and Kress' result for the
case of the diffusion equation with an absorbing inhomogeneity
embedded in an infinite homogeneous medium and derive the convergence
condition (\ref{2}). In Section~\ref{sec:discreteization} we describe
a discretization scheme for representation of operators by matrices
which is used in numerical examples of Section~\ref{sec:num}.  Here
the analytical results of Section~\ref{sec:convergence_alpha} are
verified numerically. Finally, Section~\ref{sec:discussion} contains a
discussion of obtained results.

Before proceeding with the main content of this paper, we wish to
clarify the following point. In the text below, we use the terms
``multiple scattering of diffuse waves'' and ``interaction''. We are
referring to multiple scattering of scalar solutions to the diffusion
equation from inhomogeneities in its coefficients -- not to multiple
scattering of electromagnetic waves from inhomogeneities in the
dielectric susceptibility. The first effect can be viewed as
macroscopic, and takes place on much larger scales than the second
effect. In particular, a macroscopically homogeneous medium with
constant absorption and diffusion coefficients exhibits no scattering
of diffuse waves, although the very possibility to describe the
electromagnetic energy density by the diffusion equation is based on
the assumption of strong multiple scattering of electromagnetic waves
on microscopic physical scales. Similarly, by ``interaction'' we mean
the interaction (interference and multiple scattering) of diffuse
waves scattered from macroscopic inhomogeneities.

\section{Derivation of the Born Series}
\label{sec:Born}

Propagation of light in biological tissues is commonly described by
the diffusion approximation to the radiative transport
equation~\cite{arridge_99_1,gibson_05_1}. In the case of
continuous-wave illumination, the following steady-state diffusion
equation is used:

\begin{equation}
\label{DE_1}
[-\nabla \cdot D({\bf r}) \nabla + \alpha({\bf r})]u({\bf r}) = q({\bf
  r}) \ ,
\end{equation}

\noindent
where $u$ is the energy density of the diffuse light inside the
medium, $q$ is the source function, $D = c/[3(\mu_a +
\mu_s^{\prime})]$, $\alpha = c\mu_a$ and $c$ is the average speed of
light in the medium. Further, $\mu_a$ and $\mu_s^{\prime}$ are the
absorption and reduces scattering coefficients, respectively.
Reconstruction of the functions $\mu_a({\bf r})$ and
$\mu_s^{\prime}({\bf r})$ from a set of boundary measurements is the
goal of OT.

Experiments in OT are usually performed with point sources
(plane-wave~\cite{markel_05_3} or structured~\cite{cuccia_05_1}
illumination have also been proposed). A point source can be written
as $q({\bf r}) = q_0 \delta({\bf r}-{\bf r}_s)$. Here ${\bf r}_s$ is
the source location on the boundary of the medium. A point detector
located at ${\bf r}_d$ can be shown~\cite{markel_04_4} to produce a
measurement that is proportional to the Green's function of
Eq.~(\ref{DE_1}), $G({\bf r}_d, {\bf r}_s)$, which satisfies

\begin{equation}
\label{GF}
[\nabla \cdot D ({\bf r}) \nabla - \alpha({\bf r})]G({\bf r},{\bf r}^{\prime}) = -
\delta({\bf r}-{\bf r}^{\prime}) \ . 
\end{equation}

We now decompose $\alpha({\bf r})$ and $D({\bf r})$ as constant
background values $\alpha_0$, $D_0$ and spatially-varying functions
$\delta\alpha({\bf r})$, $\delta D({\bf r})$, according to
$\alpha({\bf r}) = \alpha_0 + \delta\alpha({\bf r})$ and $D({\bf r}) =
D_0 + \delta D({\bf r})$. The background constants are chosen to be
equal to the respective values of $\alpha$ and $D$ near the medium
boundary where these coefficients are either directly measurable or
known, i.e., by immersing the sample into a matching fluid whose
optical properties are known. We then obtain the Dyson equation for
the Green's function~\cite{schotland_97_1,gonatas_95_1}, namely

\begin{equation}
\label{Dyson_int}
G({\bf r},{\bf r}^{\prime}) = G_0({\bf r},{\bf r}^{\prime}) + \int
G_0({\bf r},{\bf r}^{\prime\prime}) V({\bf r}^{\prime\prime}) G({\bf
  r}^{\prime\prime},{\bf r}^{\prime}) d^3r \ , 
\end{equation}

\noindent
where the integration is over the spatial region occupied by the
scattering medium, $G_0({\bf r},{\bf r}^{\prime})$ is the Green's
function for a homogeneous medium with $\alpha=\alpha_0$ and $D=D_0$,
i.e., it satisfies

\begin{equation}
\label{GF_0}
[D_0 \nabla^2 - \alpha_0]G_0({\bf r},{\bf r}^{\prime}) = -
\delta({\bf r} - {\bf r}^{\prime})
\end{equation}

\noindent
and appropriate boundary conditions on the scattering medium boundary,
and $V({\bf r})$ is given by

\begin{eqnarray}
\label{V_def}
& V({\bf r}) = V_\alpha({\bf r}) + V_D({\bf r}) \ , \\
\label{V_alpha_def}
& V_\alpha({\bf r}) = -\delta\alpha({\bf r})  \ ,\\
\label{V_D_def}
& V_D({\bf r}) = -{\bf p} \cdot \delta D({\bf r}) {\bf p} \ .
\end{eqnarray}

\noindent
Here we have introduced the momentum operator ${\bf p}=-i\nabla$.
Since ${\bf p}$ is Hermitian (self-adjoint), so is $V_D$. We note that
the Dyson equation (\ref{Dyson_int}) is valid for ${\bf r}$, ${\bf
  r}^{\prime}$ being inside the scattering medium or on its boundary.
In the latter case, we can replace ${\bf r}$ and ${\bf r}^{\prime}$ by
${\bf r}_d$ and ${\bf r}_s$.

In operator notation, the Dyson equation (\ref{Dyson_int}) is written
as

\begin{equation}
\label{Dyson_oper}
G = G_0 + G_0 V G \ ,
\end{equation}

\noindent
where $V = V_\alpha + V_D$ is the interaction operator. We note that
$V_\alpha$ is diagonal in the position representation and has the
matrix elements

\begin{equation}
\label{V_alpha_rs}
\langle {\bf r} \vert V_\alpha \vert {\bf r}^{\prime} \rangle = -
\delta\alpha({\bf r}) \delta({\bf r} - {\bf r}^{\prime}) \ .
\end{equation}

\noindent
However, $V_D$ has no position representation.~\footnote[3]{Of course,
  differential operators in Eq.~(\ref{DE_1}) can be approximated by
  finite differences. However, all finite difference schemes are
  non-local (involve several spatial points) and, strictly speaking,
  can not be used to define a position representation of $V_D$.} Its
matrix elements can be defined in the basis of plane waves (in ${\bf
  k}$-space).  For example, in an infinite space we can take the basis
functions to be $\vert \psi_{\bf k} \rangle$, such that $\langle {\bf
  r} \vert \psi_{\bf k} \rangle = (2\pi)^{-3/2} \exp(i {\bf k} \cdot
{\bf r})$, Then we have the following matrix elements (of both
$V_\alpha$ and $V_D$):

\begin{eqnarray}
\label{V_alpha_k}
& \langle \psi_{{\bf k}^{\prime}} \vert V_\alpha \vert \psi_{{\bf k}}
\rangle = - \delta\tilde{\alpha}({\bf k} - {\bf k}^{\prime}) \ , \\
\label{V_D_k}
& \langle \psi_{{\bf k}^{\prime}} \vert V_D \vert \psi_{{\bf k}}
\rangle = -{\bf k}^{\prime}  \cdot {\bf k} \ 
 \delta\tilde{D}({\bf k} - {\bf k}^{\prime}) \ ,
\end{eqnarray}

\noindent
where the tilde denotes three-dimensional Fourier transform with
respect to the spatial variable ${\bf r}$. The simple mathematical
structure of the above matrix elements suggests that the forward and
inverse problems are more naturally formulated in ${\bf k}$-space,
especially if the medium boundaries are translationally
invariant~\cite{markel_03_1,markel_04_4}.

The Born series is obtained by iterating (\ref{Dyson_oper}) starting
with $G=G_0$ and has the form

\begin{equation}
\label{Born_series_G}
G = G_0 + G_0 V G_0 + G_0 V G_0 V G_0 + \ldots = G_0
\sum_{k=0}^{\infty} (V G_0)^k \ .
\end{equation}

\noindent
The Born series can also be viewed as the Taylor expansion of the
formal solution to (\ref{Dyson_oper}) into a power series in $V$,

\begin{equation}
\label{G_G0_V}
G = (I - G_0 V)^{-1} G_0 = G_0 (I - V G_0)^{-1} \ ,
\end{equation}

\noindent
$I$ being the identity operator.

The derivation of the convergence condition can be obtained directly
starting from Eq.~(\ref{G_G0_V}). However, a more mathematically
elegant approach can be based on an analogous formula for the
T-matrix. In the T-matrix formalism, one writes the Dyson equation
(\ref{Dyson_oper}) as

\begin{equation}
\label{Dyson_T}
G = G_0 + G_0 T G_0 \ .
\end{equation}

\noindent
From the identity $TG_0 = VG$, we obtain $T=V G G_0^{-1}$ or,
substituting this into (\ref{G_G0_V}),

\begin{equation}
\label{T_G0_V}
T = V (I - G_0 V)^{-1} = (I - V G_0)^{-1} V \ .
\end{equation}

\noindent
The Born series for the T-matrix is

\begin{equation}
\label{Born_series_T}
T = V + V G_0 V + V G_0 V G_0 V + \ldots = \left[ \sum_{k=0}^{\infty} (V
G_0)^k \right] V \ .
\end{equation}

\noindent
Note that the series in (\ref{Born_series_G}) and
(\ref{Born_series_T}) are identical and, therefore, the convergence
conditions for the series expansions of $G$ and $T$ are also
identical.

\section{The Convergence Condition for Absorbing Inhomogeneities}
\label{sec:convergence_alpha}

The diffusion approximation is valid when $\mu_s^{\prime} \gg \mu_a$.
If, in addition, $\mu_s^{\prime}$ is constant inside the sample, then
$D({\bf r})$ is also, approximately, constant. This case is
practically important when the contrast mechanism is directly related
to absorption, but not to scattering, for instance, in imaging of
blood oxygenation levels~\cite{culver_03_3}.

In this Section, we specialize to the case $\delta D = 0$,
$\delta\alpha \neq 0$, so that $V=V_\alpha$. We say that the function
$\delta\alpha({\bf r})$ is {\em physically allowable} if
$\delta\alpha({\bf r}) \geq -\alpha_0$. In the opposite case, the
total absorption coefficient $\alpha({\bf r}) = \alpha_0 +
\delta\alpha({\bf r})$ can become negative, which physically
corresponds to an amplifying medium.

The derivations presented below are based on the assumption that for
any physically allowable $\delta\alpha$, the diffusion equation
(\ref{DE_1}) has a solution. We also use the fact that if
$\delta\alpha$ is physically allowable and satisfies $\delta\alpha
\leq \alpha_0$, then $-\delta\alpha$ is also physically allowable.
While we assume on physical grounds that Eq.~(\ref{DE_1}) has a
solution for every physically allowable $\delta\alpha$, it can not be
stated that if $\delta\alpha$ {\em is not physically allowable}, then
(\ref{DE_1}) {\em has no solutions}.  In fact, (\ref{DE_1}) can have a
steady-state solution even if the medium is amplifying in some finite
spatial region, as long as there also exists a sufficiently strong
energy sink~\footnote[5]{If $D=D_0={\rm const}$, the diffusion
  equation (\ref{DE_1}) is mathematically equivalent to the
  Schroedinger equation for a single particle of mass $m$ in the
  potential $U({\bf r})=(\hbar^2/2m)\alpha({\bf r})/D_0$. From the
  analysis presented below, it will be clear that the solution to
  (\ref{DE_1}) ceases to exists if the potential $U({\bf r})$ is deep
  enough to support at least one bound state. See
  Ref.~\cite{bushell_72_1} for a similar argument.}. For this reason,
the convergence conditions derived in
Sections~\ref{sec:convergence_alpha},\ref{sec:convergence_D} and are
sufficient but not necessary.

\subsection{Sign-Definite $\delta\alpha$.}
\label{subsec:sign_def}

We start with the simple case of a sign-definite function
$\delta\alpha({\bf r})$.  Namely, we assume that $\delta\alpha({\bf
  r})$ does not change sign within its domain (but can be zero). We
also assume that $\delta\alpha({\bf r})$ has no singularities. Then we
can write

\begin{equation}
\label{V_S_definite}
V = - \sigma S S \ ,
\end{equation}

\noindent
where $\sigma=\pm 1$ and $S$ is a non-negative definite operator,
diagonal in the position representation. The values of $\sigma$ are
$\sigma=+1$ if $\delta\alpha \geq 0$ and $\sigma=-1$ if
$\delta\alpha\leq 0$.  Then, with little algebraic manipulation, we
obtain

\begin{equation}
\label{T_symmetric}
T = -\sigma S (I + \sigma S G_0 S)^{-1} S = -\sigma S (I + \sigma
W)^{-1} S \ .
\end{equation}

\noindent
In the above formula, $W=S G_0 S$. The matrix elements of $W$ are
given by

\begin{equation}
\label{W_def}
\langle {\bf r}
\vert W \vert {\bf r}^{\prime} \rangle = \sqrt{ \vert
  \delta\alpha({\bf r}) \vert } \ G_0({\bf r}, {\bf r}^{\prime}) \ \sqrt{ \vert
  \delta\alpha({\bf r}^{\prime}) \vert } \ .
\end{equation}

\noindent
The operator $W$ can be viewed as a functional of $\delta\alpha$.  We
note the following obvious property: $W[\gamma \delta\alpha] =
\vert\gamma\vert W[\vert \delta\alpha \vert ]$, where $\gamma$ is a
constant.

$W$ is real and symmetric so that all of its eigenvalues $w_\mu$ are
real. The Born series (\ref{Born_series_T}) converges if all
eigenvalues satisfy $\vert w_\mu \vert < 1$ and diverges otherwise. We
note that the index $\mu$ that labels the eigenvalues may not be
countable, i.e., if the spectrum of $W$ is continuous.  Of course, the
eigenvalues $w_\mu$ are not computable analytically in general and the
above condition is of little practical use. However, we will employ
the following lemma to obtain conditions on $\delta\alpha$ itself:

\paragraph{Lemma 1} For any physically allowable $\delta\alpha$ that
does not change sign, $\sigma w_\mu[\delta\alpha] \neq -1$ for all
indices $\mu$.

\paragraph{Proof} For any physically allowable
$\delta\alpha$, there is a solution to the diffusion equation
(\ref{DE_1}) and, correspondingly, a T-matrix.  For the T-matrix to
exist, the operator $I+\sigma W$ in (\ref{W_def}) must be invertible.
But if $\sigma w_\mu = -1$ for at least one eigenvalue, the above
operator is not invertible.
\\

In particular, for non-negative functions $\delta\alpha$
($\sigma=+1$), $w_\mu[\delta\alpha] \neq -1$ and for non-positive
and {\em physically allowable} functions $\delta\alpha$
($\sigma=-1$), $w_\mu[\delta\alpha] \neq +1$.  If the sign of a
physically allowable $\delta\alpha$ is reversed and $-\delta\alpha$ is
still physically allowable, then $w_\mu[\delta\alpha] \neq \pm 1$.
This property holds for all physically allowable functions
$\delta\alpha$ such that $\delta\alpha \leq \alpha_0$.

We can now state two simple results that set bounds on the spectrum of
$W$.

\paragraph{Proposition 1} For any physically allowable
$\delta\alpha$ that does not change sign, $W[\delta\alpha]$ has no
negative eigenvalues.
\paragraph{Proof} Let $W[\delta\alpha]$ have an eigenvalue $w<0$. 
Choose $\gamma = 1/\vert w \vert$. Then $W[\gamma \vert \delta\alpha
\vert]$ has an eigenvalue $-1$. Since $\gamma \vert \delta\alpha
\vert$ is non-negative, this is not possible by Lemma 1.
\\

\paragraph{Proposition 2} If, in addition to the conditions of
Proposition 1, $\delta\alpha \leq \alpha_0$, then all eigenvalues of
$W[\delta\alpha]$ are less than unity.
\paragraph{Proof}  Let $W[\delta\alpha]$ have an eigenvalue $w>1$. Choose
$\gamma = -1/w$. Then $W[ \gamma \vert \delta\alpha \vert]$ has an
eigenvalue $+1$.  Since $\gamma \vert \delta\alpha \vert$ is
physically allowable and non-positive, this is not possible by Lemma
1.
\\
 
To summarize, we have found that all eigenvalues of the matrix
$W=SG_0S$ lie in the open interval $[0,1)$ for all physically
allowable functions $\delta\alpha$ that satisfy the conditions of
Proposition 2. Since $\sigma=\pm 1$, we immediately conclude that,
under the same conditions, the expansion of (\ref{T_symmetric}) into
a power series in $W$ converges. It is further straightforward to see
that this expansion is identical to (\ref{Born_series_T}) or
(\ref{Born_series_G}).  Therefore, we have established the following
condition for convergence of the Born series:

\paragraph{Theorem} The Born series for the T-matrix or the Green's function
converges if (i) $\delta\alpha$ is physically allowable, (ii) does not
change sign inside its domain, and (iii) satisfies $\delta\alpha({\bf
  r}) \leq \alpha_0$.
\\

A remarkable feature of the above condition is that it depends only on
the upper bound for $\delta\alpha$, but not on its shape or spatial
extent. Thus, for example, let $\delta\alpha({\bf r}) = A \leq
\alpha_0$ inside some region $\Omega$. The Born series will converge
independently of the shape or linear dimensions of this region.
Physically, this can be understood by considering the fact that
$G_0({\bf r},{\bf r}^{\prime})$ decays exponentially with the distance
between ${\bf r}$ and ${\bf r}^{\prime}$.  Therefore, multiple
scattering of diffuse waves on large scales is exponentially
suppressed. Instead, scattering is strong at small scales, when
$G_0({\bf r},{\bf r}^{\prime}) \propto 1/\vert {\bf r} - {\bf
  r}^{\prime} \vert$. It is this short-range interaction that may
result in a substantially nonlinear dependence of $G(V)$ or $T(V)$ on
$V$. If $\delta\alpha/\alpha_0$ is sufficiently large, even locally,
the nonlinearity may become so strong that the power series expansion
of $T(V)$ does not converge.  However, we have established that this
expansion always converges if $\delta\alpha \leq \alpha_0$.

We conclude this subsection with the following remark. Proposition 1
is stronger than is needed for the derivation of the above convergence
condition. The inequality $w_\mu > -1$ would be sufficient. In fact,
we will see below that Proposition 1 holds only for operators $W$
whose trace is infinite.  If we perform a discretization as is
explained in Section~\ref{sec:discreteization}, $W$ becomes a
finite-size matrix of zero trace. The scaling property $W[\gamma
\delta\alpha] = \vert \gamma \vert W[\vert \delta\alpha \vert]$ does
not hold for such matrices. Consequently, some of their eigenvalues
are negative.  However, they are all greater than $-1$. The proof of
this statement is very similar to the proof of Proposition 2 and is
omitted; instead, we will illustrate this fact with numerical
examples.

\subsection{Sign-Indefinite $\delta\alpha$.}
\label{subsec:sign_indef}

We will now show that the convergence condition formulated in the
previous subsection holds even if $\delta\alpha({\bf r})$ can change
sign.

Before proceeding with the proof, we set the stage for the numerical
verification of this statement in Section~\ref{sec:num} below. Since
$\delta\alpha$ is now allowed to change sign, we can no longer write
$V=-\sigma SS $ where $\sigma = \pm 1$ and $S$ is real and
non-negative definite. Instead, we can write, for example, $V = -S_c
S_c$, where $S_c$ is complex.  Analogously to (\ref{T_symmetric}), we
have

\begin{equation}
\label{T_complex_symmetric}
T = - S_c (I + S_c G_0 S_c)^{-1} S_c = -S_c (I + W_c)^{-1} S_c  \ ,
\end{equation}

\noindent
where $W_c = S_c G_0 S_c$. The matrix elements of $W_c$ are

\begin{equation}
\label{W_c_def}
\langle {\bf r}
\vert W_c \vert {\bf r}^{\prime} \rangle = \sqrt{
  \delta\alpha({\bf r}) } \ G_0({\bf r}, {\bf r}^{\prime}) \ \sqrt{ 
  \delta\alpha({\bf r}^{\prime}) } \ .
\end{equation}

\noindent
Note that $W_c$ does not depend on the choice of the square root
branch in the above formula, as long as the same branch is chosen in
both square roots.

Since $W_c$ is complex symmetric and hence non-Hermitian, its
eigenvalues are in general complex.  Therefore, placing bounds on the
eigenvalues of $W_c$ is problematic. Indeed, the analog of Lemma 1 for
Eq.~(\ref{T_complex_symmetric}) is $w_\mu \neq -1$. But this
inequality can be satisfied trivially if $w_\mu$ has an imaginary
part.  Therefore, Eq.~(\ref{T_complex_symmetric}) is not useful for
the derivation of a convergence condition. Instead, we will study
eigenvalues of $W_c$ numerically in Section~\ref{sec:num}.  Here we
will use a different representation for the T-matrix.  Namely, we can
write $V = - S \Sigma S$ where $S$ is still real and non-negative
definite but $\Sigma$ is now an operator rather than a number:
\begin{equation}
\label{Sigma_def}
\langle {\bf r}
\vert \Sigma \vert {\bf r}^{\prime} \rangle = \delta({\bf r} - {\bf
  r}^{\prime}) \left\{
\begin{array}{ll}
+1 \ , & {\rm if} \ \ \delta\alpha({\bf r}) \geq 0 \ , \\
-1 \ , & {\rm if} \ \ \delta\alpha({\bf r}) < 0 \ . \\
\end{array} \right.
\end{equation}

\noindent
Thus, we can refer to $\Sigma$ as the sign operator. Note that
$\Sigma$ and $S$ commute. After straightforward algebraic
manipulation, we obtain

\begin{equation}
\label{T_Sigma}
T = -S(\Sigma + S G_0 S)^{-1} S = -S (\Sigma + W)^{-1} S \ .
\end{equation}

\noindent
In the above equation, $W[\delta\alpha]$ is defined by (\ref{W_def})
of Section~\ref{subsec:sign_def}, but its domain has been generalized
to include functions $\delta\alpha$ that can change sign.  Still,
since $W[\delta\alpha] = W[\vert \delta\alpha \vert]$, and from the
results of previous subsection, we know that the eigenvalues $w_\mu$
of $W$ lie in the interval $[0,1)$, as long as $\vert \delta\alpha
\vert \leq \alpha_0$.  Therefore, $\vert\vert W \vert\vert < 1$,
where $\vert\vert \cdot \vert\vert$ is the operator norm defined
here as $\vert\vert W \vert\vert = \sup [\langle \psi \vert W \vert
\psi \rangle / \langle \psi \vert \psi \rangle]$.  On the other hand,
from the obvious relation $\Sigma^2 = I$, we find that $\vert\vert
\Sigma \vert\vert = 1$. We then write

\begin{equation}
\label{Sigma_plus_W}
(\Sigma + W)^{-1} = [\Sigma(I + \Sigma W)]^{-1} = (I + \Sigma W)^{-1}
\Sigma \ .
\end{equation}

\noindent
The Born series is obtained by expanding

\begin{equation}
\label{Sigma_plus_W_exp}
(I + \Sigma W)^{-1} = \sum_{k=0}^{\infty} (-\Sigma W)^k \ .
\end{equation}

\noindent
From the operator norm inequality $\vert\vert A B \vert\vert_p \leq
\vert\vert A \vert\vert_p \cdot \vert\vert B \vert\vert_p$, we immediately
obtain $\vert\vert \Sigma W \vert\vert < 1$, which is a sufficient
condition for convergence of the series (\ref{Sigma_plus_W_exp}). This
completes the proof that the convergence condition of the previous
subsection applies to functions $\delta\alpha({\bf r})$ that can
change sign.

\section{The Convergence Condition for Scattering Inhomogeneities}
\label{sec:convergence_D}

If $\mu_a={\rm const}$ while $\mu_s^{\prime}$ varies, the system is
characterized by a scattering inhomo\-geneity. We then have
$\delta\alpha=0$, $\delta D \neq 0$.  Obviously, the physically
allowable values of $\delta D$ satisfy $\delta D \geq - D_0$.
However, the physical interpretation of what happens if we do allow
$D({\bf r})$ to become negative is somewhat different. If the source
function of Eq.~(\ref{DE_1}) is zero in the spatial region where $D$
is negative, than the interpretation is that the medium in that region
is amplifying, similar to the case of absorbing inhomogeneities. But
if $D$ is negative in a region where the source is nonzero, then, in
addition to having amplifying medium, the source of energy is turned
into a sink.

We now restrict consideration to a physically allowable $\delta D$ and
state that the convergence condition of
Section~\ref{sec:convergence_alpha} applies to scattering
inhomogeneities with the substitution $\alpha_0 \rightarrow D_0$ and
$\delta\alpha \rightarrow \delta D$. The proof of this statement is
analogous to the proof given in Section~\ref{sec:convergence_alpha}
and will be only briefly sketched.

For a general physically allowable $\delta D$, the interaction
operator can be written as $V = V_D = - {\bf p} \cdot S \Sigma S {\bf
  p}$ and the symmetric expression for the T-matrix, analogous to
(\ref{T_Sigma}), is

\begin{equation}
\label{T_S_diffusion}
T = -{\bf p} \cdot S \left[\Sigma + S {\bf p} G_0 {\bf p} \cdot S
\right]^{-1} S {\bf p} \ .
\end{equation}

\noindent
The operator $W = S {\bf p} G_0 {\bf p} \cdot S$ is complex but
Hermitian, so that all of its eigenvalues are strictly real. By
considering the special cases of sign-definite $\delta D$ when $\Sigma
= \pm I$, we obtain bounds on the eigenvalues of $W$ in complete
analogy with Section~\ref{subsec:sign_def}. More specifically, the
eigenvalues of $W$ all lie in the open interval $[0,1)$, as long as
$\delta D \leq D_0$.  We then find that the operator norm of $W$ is
less than unity while it is exactly unity for $\Sigma$, and,
consequently, expansion of (\ref{T_S_diffusion}) into a power series
converges.

\section{Generalization of Colton and Kress' Result}
\label{sec:colton}

Further insight into the convergence properties of the Born series and
the strength of nonlinearity can be gained by considering the argument
similar to the one used by Colton and Kress in the proof of Theorem
8.4 of Ref.~\cite{colton_book_98}. The argument is based on a direct
estimation of the norm $\vert\vert VG_0 \vert\vert_\infty$ of the
operator $VG_0$ that appears in the series (\ref{Born_series_G}) or
(\ref{Born_series_T}). The (necessary and sufficient) convergence
condition for the Born series is $\vert\vert VG_0
\vert\vert_\infty<1$. Of course, estimation of this norm is possible
only if $G_0$ is known analytically. For a medium with boundaries,
$G_0$ can only be computed numerically. Therefore, we will consider
below the simple case of free space, so that

\begin{equation}
\label{GF_free}
G_0({\bf r}, {\bf r}^{\prime}) = G_F({\bf r}, {\bf r}^{\prime}) =
\frac{\exp(-k_d \vert {\bf r} - {\bf r}^{\prime} \vert) }{4\pi D_0 \vert
  {\bf r} - {\bf r}^{\prime} \vert} \ ,
\end{equation}

\noindent
where $k_d = \sqrt{\alpha_0/D_0}$ is the diffuse wave number. However,
note that the influence of boundaries can be exponentially small, as
is discussed in Section~\ref{sec:discreteization} below.

Next, we specialize to the case of absorbing inhomogeneities,
$V=V_\alpha$, where $V_\alpha$ is defined by (\ref{V_alpha_def}).
Assuming that $\delta\alpha({\bf r})=0$ if ${\bf r}$ is outside of a
sphere of radius $a$, we have

\begin{equation}
\label{VG0_norm_inf}
\vert\vert V G_0 \vert\vert_\infty \leq \sup_{\vert{\bf r}\vert \leq a} \left( \vert \delta\alpha({\bf r})\vert \right) \sup_{\vert {\bf r}\vert \leq a} \left( \vert I({\bf
  r}) \vert \right) \ ,
\end{equation}

\noindent
where

\begin{equation}
\label{I_r_def}
I({\bf r}) = \int_{r^{\prime}<a} G_F({\bf r}, {\bf r}^{\prime}) d^3
r^{\prime} \ .
\end{equation}

\noindent
The above integral can be easily evaluated to yield

\begin{equation}
\label{I_r_eval}
I({\bf r}) = \frac{1}{D_0 k_d^2} \left[ 1 - (1 + k_d a)\exp(-k_d a)
  \frac{\exp(k_d r) - \exp(-k_d r)}{2 k_d r} \right] \ .
\end{equation}

\noindent
Obviously, the maximum of the above function is at the center of the
ball, so that

\begin{equation}
\label{maxI_f}
\sup_{\vert {\bf r}\vert \leq a}\left( \vert I({\bf r}) \vert \right) = \frac{1}{D_0
  k_d^2}f(k_d a) \ , \ \ f(x) = 1 - (1+x)\exp(-x) \ . 
\end{equation}

\noindent
We then immediately arrive at the (sufficient) convergence condition
(\ref{2}) for $\delta\alpha$.

We now examine the two limiting cases $k_da \rightarrow 0$ and $k_da
\rightarrow \infty$. In the first case, we use $f(x) \approx x^2/2$
for small $x$ and recover Colton and Kress' convergence condition
$\delta\alpha < 2\alpha_0/(k_d a)^2$. In the second case, the domain
of $\delta\alpha$ is not restricted and we recover the result of
Section~\ref{sec:convergence_alpha}, namely, $\delta\alpha < \alpha_0$
with the only difference that we now have a strict inequality. The
independence of the latter result on $k_da$ is specific to the
diffusion equation and results from the exponential decay of diffuse
waves. Indeed, we have $\lim_{k_d a \rightarrow \infty} f(k_d a) = 1$.
However, if we perform the analytic continuation $k_d \rightarrow ik$,
the corresponding limit is $\lim_{k a \rightarrow \infty} \vert f(ika)
\vert = ka$ and the convergence condition becomes $\eta < 1/ka$ (we
have replaced here $\delta\alpha/\alpha_0$ by its counterpart $\eta$).
This fact illustrates the crucial difference in convergence properties
of the Born series for propagating and diffuse waves.

\section{Discretization}
\label{sec:discreteization}

In any numerical simulations, the operators $G_0$, $V$ must be
discretized and truncated using some appropriate basis.  Here we
restrict our attention to absorptive inhomogeneities so that
$V=V_\alpha$ and use the basis of cubic voxels. We note that the same
discretization method can not be applied to $V_D$ because, as was
mentioned in Section~\ref{sec:Born}, $V_D$ has no position
representation.

The discretization method described below is analogous to the
so-called discrete-dipole approximation~\cite{purcell_73_1} that has
been widely used in electromagnetic scattering by nonspherical
particles~\cite{draine_88_1,draine_00_1}. We seek to discretize the
integral equation (\ref{Dyson_int}) in a basis of cubic voxels.
Instead of working directly with (\ref{Dyson_int}), it is more
convenient to first write the Lippmann-Schwinger integral equation for
the field $u$ itself. Let $u({\bf r}) = \int G({\bf r},{\bf
  r}^{\prime}) q({\bf r}^{\prime}) d^3r^{\prime}$ and $u^{\rm
  inc}({\bf r}) = \int G_0({\bf r},{\bf r}^{\prime}) q({\bf
  r}^{\prime}) d^3r^{\prime}$. Here $u^{\rm inc}({\bf r})$ is the
incident field, i.e., the field that would exist in the absence of
inhomogeneities. Using $V=V_\alpha$, we obtain the following integral
equation for $u({\bf r})$:

\begin{equation}
\label{Lippman}
u({\bf r}) = u^{\rm inc}({\bf r}) - \int G_0 ({\bf r},{\bf
  r}^{\prime}) \delta\alpha({\bf r}^{\prime}) u({\bf r}^{\prime})
d^3r^{\prime} \ .
\end{equation}

\noindent
We then break up the sample into cubes $C_n$ of side $h$, volume
$v=h^3$, and denote the center of each cube by ${\bf r}_n$. The field
$u({\bf r})$ is approximated by a set of discrete values $u_n=u({\bf
  r}_n)$.  Setting ${\bf r} = {\bf r}_n$ in (\ref{Lippman}) and
representing the volume integral as a sum of integrals over each
voxel, we obtain

\begin{equation}
\label{Lippmann_1}
u_n = u^{\rm inc}_n - \sum_m \int_{C_m} \ 
G_0 ({\bf r}_n, {\bf r}) \delta\alpha({\bf r}) u({\bf r}) d^3 r \ ,
\end{equation}

\noindent
where $u^{\rm inc}_n = u^{\rm inc}({\bf r}_n)$. The above equation is,
so far, exact. We now introduce several approximations. First, we
replace $\delta\alpha({\bf r}) u({\bf r})$ in the integrand of
Eq.~(\ref{Lippmann_1}) by $\delta\alpha_m u_m$, where $\delta\alpha_n =
\delta\alpha({\bf r}_n)$. Second, in all terms with $m\neq n$, we
replace $G_0 ({\bf r}_n, {\bf r})$ by $G_0 ({\bf r}_n, {\bf r}_m)$.
We then have

\begin{eqnarray}
\label{Lippmann_2}
& u_n = u^{\rm inc}_n - \sum_{m \neq n} G_0({\bf r}_n, {\bf r}_m)
v \delta\alpha_m u_m 
- Q_n \delta\alpha_n u_n \ , \\
\label{Q_def}
& Q_n = \int_{C_n} G_0 ({\bf r}_n, {\bf r}) d^3
r \ . 
\end{eqnarray}

\noindent
Note that the term with $m=n$ has been treated separately because the
homogeneous medium Green's function $G_0({\bf r},{\bf r}^{\prime})$
has a singularity at ${\bf r} = {\bf r}^{\prime}$. The singularity is
integrable and the quantity $Q_n$ is well defined. However, the
computation of $Q_n$ is complicated due to the following two factors.
First, $G_0 ({\bf r}, {\bf r}^{\prime})$ depends on the shape of
boundaries and on the extrapolation distance $\ell$ in a complicated
way and is not computable analytically in general. Second, the
integration in (\ref{Q_def}) is over a cubic volume, while the
asymptotic $\lim_{{\bf r} \rightarrow {\bf r}^{\prime}} [G_0 ({\bf r},
{\bf r}^{\prime})] \propto 1/\vert {\bf r} - {\bf r}^{\prime}\vert$
has spherical symmetry. The first difficulty is resolved by noting
that $G_0$ is a sum of the Green's function in an infinite homogeneous
space $G_F$ and a contribution due to the boundaries $G_B$:

\begin{equation}
\label{G0_GF_GB}
G_0({\bf r}, {\bf r}^{\prime}) = G_F({\bf r}, {\bf r}^{\prime}) +
G_B({\bf r}, {\bf r}^{\prime}) \ ,
\end{equation}

\noindent
where $G_F$ is given by (\ref{GF_free}). Accordingly, we can write
$Q_n$ as a sum of two contributions, $Q_F$ and $Q_{Bn}$. Note that the
$Q_F$ is independent of the index $n$ because the Green's function in
an infinite homogeneous space is translationally invariant. The term
$Q_{Bn}$ can depend on $n$ because boundaries break translational
invariance, so that the integral in (\ref{Q_def}) can depend on ${\bf
  r}_n$. However, we will argue that $Q_{Bn}$ is a small correction to
$Q_F$. Indeed, $G_B({\bf r}_n, {\bf r})$ can be written as a surface
integral taken over the medium boundaries and has no singularity at
${\bf r} = {\bf r}_n$.  We estimate that $Q_{Bn}/Q_F \sim
(h/L_n)\exp(-k_d L_n)$, where $L_n$ is the characteristic distance
from the point ${\bf r}_n$ to the medium boundary. We assume that all
inhomogeneities are localized in a spatial region which is
sufficiently far from the medium boundaries.  Then the ratio
$Q_{Bn}/Q_F$ is at least of the order of $h/L_n$; if, in addition,
$k_d L \gg 1$, this ratio is exponentially small.  Therefore, we will
neglect the term $Q_{Bn}$. The second difficulty is resolved by
replacing the integration over the cube $C_n$ by integration over a
sphere of equivalent volume centered at ${\bf r}_n$.  The radius of
this sphere is $R_{\rm eq} = (3/4\pi)^{1/3} h$. With these two
approximations, and using (\ref{I_r_def}), (\ref{I_r_eval}), we have

\begin{equation}
\label{Q_computed}
Q_n = Q_F = \frac{1}{k_d^2 D_0} f(k_d R_{\rm eq}) \ ,
\end{equation}

\noindent
where $f(x)$ is defined in (\ref{maxI_f}).  Note that for small $x$,
$Q_F \approx R_{\rm eq}^2/2 D_0$.

Having computed $Q_F$, we can write a self-consistent ``coupled-dipole
equation'' which is a discrete approximation to the integral equation
(\ref{Lippmann_1}).~\footnote[8]{In the case of the scalar field
  $u({\bf r})$, a more appropriate term is ``coupled-monopole
  equation'' since the quantities $d_n$ are, in fact, monopoles. We,
  however, adhere to the terminology used in electromagnetic
  scattering theory.} We define ``dipole moments'' $d_n = - v u_n
\delta\alpha_n$, and, after some rearrangement of (\ref{Lippmann_2}),
obtain

\begin{eqnarray}
\label{CDE}
& d_n = \chi_n \left[ u^{\rm inc}_n + \sum_{m \neq n} G_0({\bf r}_n,
  {\bf r}_m) d_m \right] \ , \\
\label{chi_def}
& \chi_n = - \frac{v \delta\alpha_n}{1 + Q_F \delta\alpha_n} \ .
\end{eqnarray}

\noindent
In the above equation, $\chi_n$ plays the role of polarizability of
the $n$-th dipole. In the absence of interaction, $d_n = \chi_n u^{\rm
  inc}_n$. Note that the polarizability depends on $\delta\alpha_n$
nonlinearly due to the presence of the term $Q_F\delta\alpha_n$ in the
denominator. A nonzero value of $Q_F$ can be viewed as a result of
interaction of the $n$-th dipole with itself and therefore can be
referred to as the dipole self-energy. The physical effect of
self-interaction is to limit the polarizability.  Thus, the maximum
(in absolute value) polarizability obtained in the limit
$\delta\alpha_n \rightarrow \infty$ is $-v/Q_F$. We note that in the
limit $k_d R_{\rm eq} \rightarrow 0$, $Q_F\delta\alpha_n \ll 1$.  In
practice, the term $Q_F \delta\alpha_n$ can be small but not zero and
should be accounted for.

We now return to operator notation. Let $\vert d \rangle$ be an
$N$-dimensional vector of dipole moments $d_n$, $n=1,\ldots,N$, where
$N$ is the total number of voxels. Similarly, we define the
$N$-dimensional vector $\vert u^{\rm inc} \rangle$.  We then have

\begin{equation}
\label{CDE_oper}
\vert d \rangle = V_\alpha \left[ \vert u^{\rm inc}\rangle +
  G_0^{\rm VV} \vert d \rangle \right] \ .
\end{equation}

\noindent
Here $V_\alpha$ and $G_0^{VV}$ are $N\times N$-matrices with elements

\begin{eqnarray}
\label{V_matrix_def}
& \langle n \vert V_\alpha \vert m \rangle  = \chi_n \delta_{nm} \ , \\
\label{GVV_matrix_def}
& \langle n \vert G_0^{\rm VV} \vert m \rangle  = (1 - \delta_{nm})
G_0({\bf r}_n, {\bf r}_m) \ . 
\end{eqnarray}

\noindent
In the above formula, the superscript ``${\rm VV}$'' is an
abbreviation for ``volume-to-volume'' and is used to emphasize that
${\bf r}_n$ and ${\bf r}_m$ are inside the discretized region. The
formal solution to (\ref{CDE_oper}) is

\begin{equation}
\label{CDE_matrix_solution}
\vert d \rangle = \left( I - V_\alpha G_0^{\rm VV} \right)^{-1}
V_\alpha \ .
\end{equation}

\noindent
If there are $N_s$ discrete sources located at the points ${\bf r}_{sk}$
($k=1,\ldots,N_s$) and $N_d$ discrete detectors at points ${\bf
  r}_{dl}$ ($l=1,\ldots,N_d$), we can write within the same precision
as was used to discretize Eq.~(\ref{Lippmann_1}):

\begin{equation}
\label{Dyson_matrix}
G^{\rm DS} = G_0^{\rm DS} + G_0^{\rm DV} \left( I - V_\alpha G_0^{\rm
  VV} \right)^{-1} V_\alpha G_0^{\rm VS}  \ .
\end{equation}

\noindent
where the matrices $G^{\rm DS}$, $G_0^{\rm DS}$, $G_0^{\rm DV}$ and
$G_0^{\rm VS}$ have the following elements:

\begin{eqnarray}
\label{G_DS_def}
& \langle l \vert G^{\rm DS} \vert k \rangle = G({\bf r}_{dl},
{\bf r}_{sk}) \ , \\
\label{G0_DS_def}
& \langle l \vert G_0^{\rm DS} \vert k \rangle = G_0({\bf r}_{dl},
{\bf r}_{sk}) \ , \\
\label{G0_DV_def}
& \langle l \vert G_0^{\rm DV} \vert n \rangle = G_0({\bf r}_{dl},
{\bf r}_n) \ , \\
\label{G0_VS_def}
& \langle n \vert G_0^{\rm VS} \vert k \rangle = G_0({\bf r}_n,
{\bf r}_{sk}) \ .
\end{eqnarray}

\noindent
Thus, $G^{\rm DS}$ and $G_0^{\rm DS}$ are matrices of size $N_d \times
N_s$, $G_0^{\rm DV}$ is of size $N_d \times N$ and $G_0^{\rm VS}$ is
of the size $N \times N_s$. The superscripts ``${\rm VS}$'' and
``${\rm DV}$'' stand for ``source-to-volume'' and
``volume-to-detector'', respectively.

Eq.~(\ref{Dyson_matrix}) is a discrete approximation to
(\ref{Dyson_T}). We can identify

\begin{equation}
\label{T_discrete}
T = (I - V_\alpha G_0^{\rm VV})^{-1} V_\alpha
\end{equation}

\noindent
as the discrete approximation to the T-matrix while $V_\alpha$ and
$G_0^{\rm VV}$ as discrete $N$-dimensional approximations to the
operators $V_\alpha$ and $G_0$ that were considered in
Sections~\ref{sec:Born},\ref{sec:convergence_alpha}. We can further
define the square root of $V_\alpha$. For example, if $\delta\alpha_n$
are sign-definite, we write $V_\alpha = - \sigma S S$, where $S$ is a
diagonal matrix with the elements $\vert \chi_n \vert^{1/2}$.  Then
the T-matrix is written in the symmetric form (\ref{T_symmetric}) with
$W=S G_0^{\rm VV} S$. In the case of sign-indefinite $\delta\alpha_n$,
we write the T-matrix in the form (\ref{T_complex_symmetric}) with
$W_c = S_c G_0^{\rm VV} S_c$ and $V_\alpha = - S_c S_c$ (see
Section~\ref{subsec:sign_indef}).

The T-matrix can be computed by direct inversion of $I - V_\alpha
G_0^{\rm VV}$. This problem is well posed and has computational
complexity $O(N^3)$. It should be stressed that computation of the
T-matrix is completely independent of the sources and detectors and
only requires knowledge of $\delta\alpha({\bf r})$ and the unperturbed
Green's function $G_0({\bf r}, {\bf r}^{\prime})$. Once the T-matrix
is found, the signal for any source-detector arrangement can be
computed using (\ref{Dyson_matrix}) by direct matrix multiplication,
an operation that can be performed with computational complexity
$O[N^2 \min(N_d,N_s) + N N_d N_s]$. In a situation when the number
of measurements is approximately equal to the number of unknowns,
e.g., $N \sim N_s N_d$, the complexity of matrix multiplication is
negligible compared to the complexity of computing the T-matrix.

The T-matrix approach to solving the forward problem has several
advantages compared to finite differences or finite elements methods.
First, only the spatial regions where inhomogeneities are supported
need to be discretized. In this sense, the method is somewhat
analogous to methods involving adaptive mesh generation. Second, once
the T-matrix is computed, the measurable signal can be easily found
for an arbitrary configuration of sources and detectors. However,
unlike the finite difference and finite elements methods, the T-matrix
method requires knowledge of $G_0({\bf r}, {\bf r}^{\prime})$ which
satisfies the proper boundary conditions. We note that $G_0$ can be
found analytically for simple geometries or, in more complex cases, it
can be computed numerically {\em once}, e.g., by finite differences or
the finite-element method.

We conclude this section by noting that the discretized matrices $W$
and $W_c$ have zero trace, unlike their continuous counterparts whose
traces are infinite. This is due to the renormalization procedure that
was employed to remove the singularity of $G_0({\bf r}, {\bf
  r}^{\prime})$.  Correspondingly, the sum of all eigenvalues of $W$
or $W_c$ is zero. Some of the eigenvalues of $W$ are necessarily
negative. In practice, we will see that $W$ has many negative
eigenvalues of very small absolute value and a much smaller number of
positive eigenvalues. When $\delta\alpha_n \leq \alpha_0$, all
eigenvalues are located in the unit circle.

\section{Numerical Examples}
\label{sec:num}

We now illustrate the theoretical results of
Section~\ref{sec:convergence_alpha} with numerical examples using the
discretization scheme of Section~\ref{sec:discreteization}. All
simulations have been performed in an infinite space, so that
$G_0({\bf r}, {\bf r}^{\prime}) = G_F({\bf r}, {\bf r}^{\prime})$,
where $G_F$ is given by (\ref{GF_free}). Physically, this corresponds
to sources, detectors and the sample being immersed into an infinite
homogeneous scattering medium. However, even if the sources and
detectors are placed on the boundary (a diffuse-nondiffuse interface),
the replacement of $G_0$ by $G_F$ can be a reasonably accurate
approximation if the boundaries are sufficiently far from the
discretized region. Indeed, as was discussed in
Section~\ref{sec:discreteization}, $G_0$ can be written as a sum of
$G_F$ and $G_B$, where the boundary contribution $G_B$ has no
singularities when both of its argument are inside the medium but not
on the medium boundary. Because $G_F({\bf r},{\bf r}^{\prime})$ has a
singularity at ${\bf r} = {\bf r}^{\prime}$, it dominates $G_B$ at
small scales. Since the large-scale interaction is suppressed due to
the exponential decay of diffuse waves, the input of boundaries is
relatively insignificant for the computation of the T-matrix. However,
computation of the {\em data function} (the measurable signal)
according to the formula (\ref{Dyson_matrix}) can depend on boundary
conditions very strongly.  This is because elements of the matrices
$G_0^{\rm DV}$ and $G_0^{\rm VS}$ are the Green's functions $G_0({\bf
  r}_d, {\bf r})$ and $G_0({\bf r}, {\bf r}_s)$ where ${\bf r}_d$ and
${\bf r}_s$ are located on the medium boundary.

For the specific choice $G_0 = G_F$, the T-matrix depends
parametrically on $k_d^2 = \alpha_0 / D_0$ but not on $\alpha_0$ and
$D_0$ separately. The same is true for $W$ and $W_c$. The quantity
$k_d$ is known as the diffuse wave number and $\lambda_d = 2\pi/k_d$
as the diffuse wavelength; it gives the inverse scale on which diffuse
waves exponentially decay. In all numerical examples shown below,
$\lambda_d$ sets the physical scale of the problem. The discretization
step $h$ is not a physical scale; it merely characterizes the
precision to which we approximate the continuous field $u({\bf r})$ by
a set of discrete values $u_n$.

In the numerical simulations shown below, we have used LAPACK
subroutines implemented in Intel's MKL library. In particular, we have used the
routines DSYEVD and ZGEEV for diagonalization of real matrices $W$ and
complex symmetric matrices $W_c$, respectively. The computation time
(on an 4$\times$1.6\ GHz Itanium-II HP rx4640 server) scaled
approximately as $0.5(N/1000)^3 {\rm sec}$ for SYEVD and $12(N/1000)^3
{\rm sec}$ for ZGEEV.  We have also employed the Rayleigh quotient to
compute the maximum eigenvalue of the real matrix $W$. This method is
quite reliable and can be used to find the maximum eigenvalue of
matrices with $N\sim 70,000$ in approximately one minute (once the
matrix $G_0^{\rm VV}$ is computed, which can take several additional
minutes).

Although we show no directly relevant data, it is interesting to
comment on the efficiency of computing the T-matrix by direct
inversion of the matrix $A = I - V_\alpha G_0^{\rm VV}$ according to
(\ref{T_discrete}). In the case of sign-definite $\delta\alpha$,
factorization and subsequent inversion of $A$ by the routines DPOTRF
and DPOTRI is performed in approximately $0.14 (N/1000)^3 {\rm sec}$.
For sign-indefinite $\delta\alpha$, the routines DGETRF and DGETRI
were employed with a computational time of $0.19 (N/1000)^3 {\rm
  sec}$.  Thus, computation of the T-matrix may be a highly efficient
method of solving the forward problem of OT and can be applicable for
discretization involving up to $\sim 10^4$ voxels. We stress that only
the spatial regions that support inhomogeneities must be discretized.
The computational disadvantage of the T-matrix approach is that the
matrices $G_0^{\rm VV}$ and $A$ are dense and require large storage
and fast access to memory.

\begin{figure}
  \centerline{\input{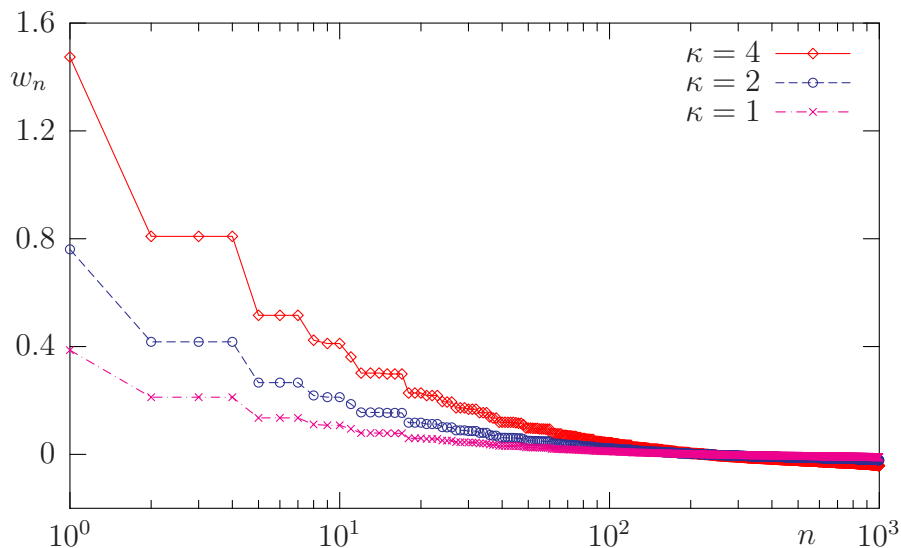}}
\caption{\label{fig:one_cube} Eigenvalues $w_n$, in descending
  order, vs the eigenvalue number $n$, for an absorbing inhomogeneity
  of cubic shape of size $H=\lambda_d/2$ and various levels of
  contrast, $\kappa$. The target is discretized by $10^3$ cubic voxels
  of size $h=\lambda_d/20$.}
\end{figure}

\subsection{Sign-Definite Case}
\label{subsec:num_sign_def}

We start with the case when $\delta\alpha({\bf r})$ does not change
sign. Namely, we compute the real symmetric matrix $W$ and find its
eigenvalues for several shapes of $\delta\alpha({\bf r})$.

The first example is an absorbing inhomogeneity (``target'') which has
the shape of a single cube with side $H=\lambda_d/2$. It was assumed
that $\delta\alpha({\bf r}) = \kappa \alpha_0$ inside the cube and is
zero outside.  The target was approximated by $10^3$ cubic voxels of
volume $h^3$. For this discretization, $h=\lambda_d/20$, $k_d R_{\rm
  eq} = 0.195$ and $Q_F\alpha_0 = 0.053$. The contrast $\kappa$ was
varied from $1$ to $4$. The eigenvalues of $W$ are shown in
Fig.~\ref{fig:one_cube}. Note that for the minimum physically
allowable contrast $\kappa=-1$, the eigenvalues differ from the case
$\kappa=1$ only very slightly due to the reversal of sign of the term
$Q_F \delta\alpha_n$ in the denominator of (\ref{chi_def}) (data not
shown). It can be seen that all eigenvalues satisfy $w_n <1$ for
$\kappa=1$ with a large margin. Obviously, the eigenvalues are even
smaller for $\kappa < 1$.

\begin{figure}
  \centerline{\input{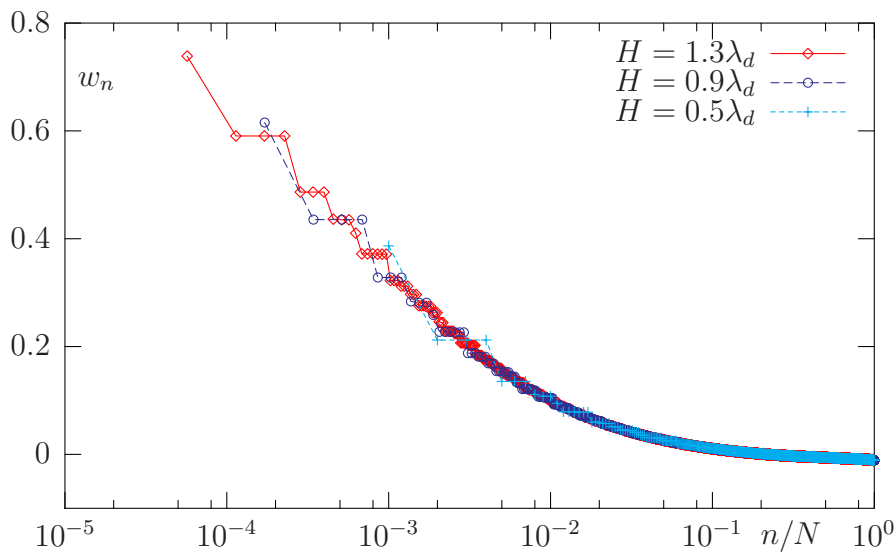}}
\caption{\label{fig:cube_H} Eigenvalues $w_n$, in descending
  order, vs the {\em relative} eigenvalue number, $n/N$, where $N$ is
  the size of the T-matrix, for an absorbing inhomogeneity of cubic
  shape, contrast $\kappa=1$, and various side length $H$. The
  discretization step is $h=\lambda_d/20$.}
\end{figure}

\begin{figure}
  \centerline{\input{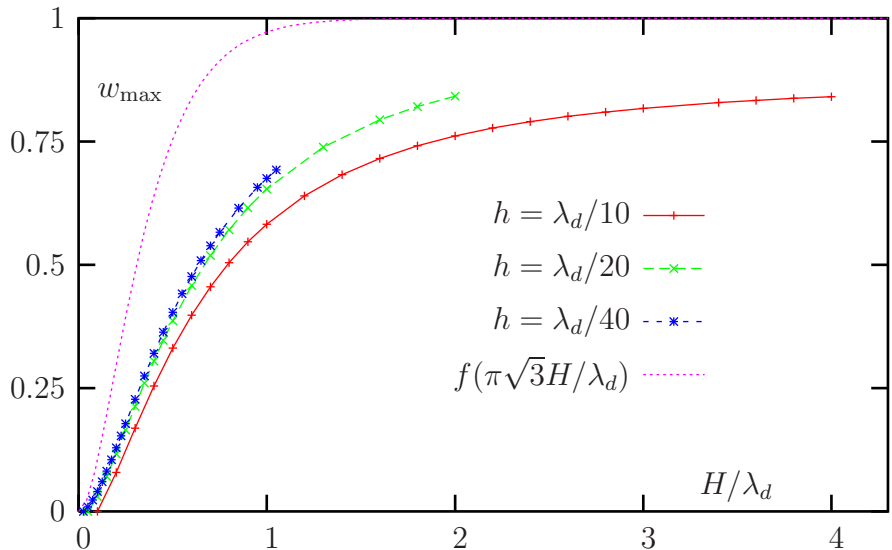}}
\caption{\label{fig:wmax} Maximum eigenvalue of $W$, $w_{\rm
    max}$, for a cubic target of contrast $\kappa=1$ as a function the
  cube size $H$ (relative to the diffuse wavelength $\lambda_d$) for
  different discretization.}
\end{figure}

Next, we fix the contrast at $\kappa=1$ and study the dependence of
eigenvalues on the size of the cubic target, $H$. In
Fig.~\ref{fig:cube_H}, we plot eigenvalues for cubes of varying sizes
$H$ while the discretization step is fixed at $h=\lambda_d/20$. It can
be seen that the maximum eigenvalue $w_{\rm max}$ (the one with the
lowest relative number) increases with the cube size but, for the set
of parameters used, does not exceed unity. To study the behavior of
$w_{\rm max}$ in a broader range of parameters, we have used the
Rayleigh quotient for various cube sizes and three different voxel
sizes (see Fig.~\ref{fig:wmax}). The Rayleigh method is well suited
for computing $w_{\rm max}$ because of the large gap between the first
two eigenvalues. The size of the cube was limited (depending on
discretization) by the computational restriction on $N$. The maximum
value of $N$ used was $N=74,088$.  Approximately one fourth of all
data points were verified by full diagonalization, with very good
agreement. It follows from Fig.~\ref{fig:wmax} that $w_{\rm max}$ does
not exceed unity for a very broad range of parameters. The curves
$w_{\rm max}(H/\lambda_d)$ approach unity from below but appear to be
unlikely to cross it. Note that inhomogeneities of sizes significantly
larger than those used in Fig.~\ref{fig:wmax} are rarely, if ever,
encountered in OT experiments since the typical value of $\lambda_d$
in biological tissues is $5{\rm cm}$. The visible difference between
curves with $h=\lambda_d/10$ and $h=\lambda_d/20$ is due to the
presence of the $h$-dependent self-energy $Q_F \delta\alpha_n$ in the
denominator of (\ref{chi_def}). This term is comparable to unity for
$h=\lambda_d/10$ but is already small for $h=\lambda_d/20$. Therefore,
the difference between the $h=\lambda_d/40$ and the $h=\lambda_d/20$
curves is insignificant. Note that we expect that discretization with
$h=\lambda_d/10$ is too rough to produce accurate results.  However,
the difference (or the lack of it) between the curves $w_{\rm
  max}(H/\lambda_d)$ with different $h/\lambda_d$ can not be used {\em
  per se} to verify convergence of the T-matrix with $h$.

Since we have performed numerical simulations in infinite space, it is
possible to compare $w_{\rm max}(H/\lambda_d)$ with the result that
can be inferred from the convergence condition (\ref{2}). To this end,
we note the following. The data for Fig.~\ref{fig:wmax} were computed
for a cube of contrast $\kappa=1$. If we increase the contrast by the
factor $\gamma$, the Born series will still converge as long as
$\gamma w_{\rm max} < 1$, or, equivalently, $\delta\alpha/\alpha_0 <
1/w_{\rm max}$. On the other hand, the convergence condition (\ref{2})
has the form $\delta\alpha/\alpha_0 < 1/f(k_d a)$, where $f(x)$ is
defined by (\ref{maxI_f}) and $a$ is the radius of the smallest sphere
that circumscribes the cube of side $H$, namely, $a=\sqrt{3}H/2$. For
these two conditions to be consistent, we must have $w_{\rm
  max}(H/\lambda_d) < f(\pi\sqrt{3}H/\lambda_d)$. The latter function
is shown as a dotted line in Fig.~\ref{fig:wmax}.

\begin{figure}
  \centerline{\input{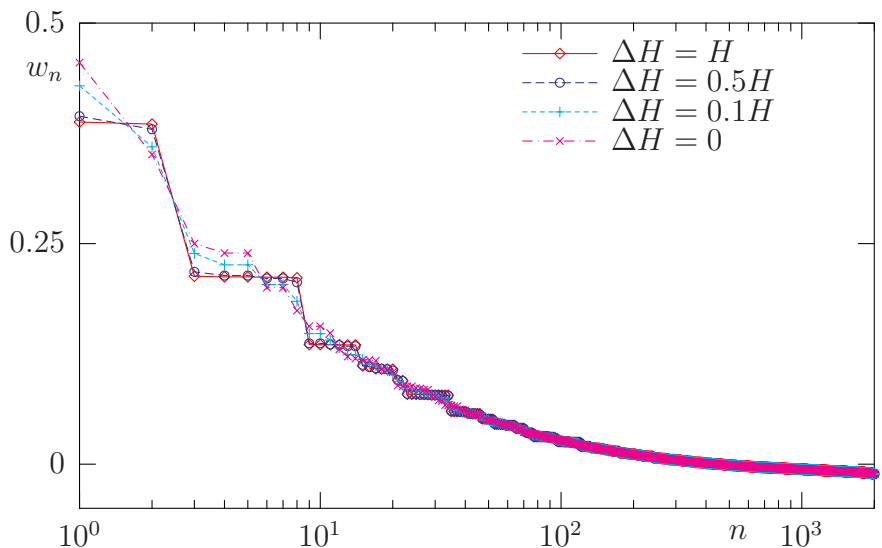}}
\caption{\label{fig:two_cubes} All eigenvalues $w_n$ of the matrix $W$ (in
  descending order) vs the eigenvalue number $n$ for an absorbing
  inhomogeneity of contrast $\kappa=1$ in the shape of two equivalent
  cubes of side $H=\lambda_d/2$ placed side-by-side and separated by
  the surface-to-surface distance $\Delta H$. Each cube was
  discretized using $h=\lambda_d/20$ ($10^3$ voxels per cube).}
\end{figure}

Next, we consider the effects of multiple scattering of diffuse waves
between two spatially separated absorbing inhomogeneities. To this
end, we plot the spectrum of eigenvalues of $W$ for two equivalent
cubic targets of contrast $\kappa=1$ and side $H=\lambda_d/2$, placed
side-by-side and separated by the surface-to-surface distance $\Delta
H$. The targets were discretized using $h=\lambda_d/20$, so that each
cube was approximated by $10^3$ voxels. The results are shown in
Fig.~\ref{fig:two_cubes}. When the cubes are sufficiently far apart
($\Delta H = H$), the interaction is weak and each eigenstate is
doubly degenerate (this is in addition to the triple degeneracy of
some eigenvalues which is due to the cubic symmetry).  When the cubes
approach, the degeneracy is broken by interaction.  However, the
effect of interaction is weak even when the two cubes approach each
other very closely. At $\Delta H= 0$, the two cubes merge and form a
single parallelepiped. At this point, the maximum eigenvalue is
increased only by $17\%$ compared to the noninteracting limit. The
weak interaction of spatially separated inhomogeneities is consistent
with the idea of exponentially suppressed long-range interaction which
was discussed in Section~\ref{subsec:sign_def}.

\subsection{Sign-Indefinite Case}
\label{subsec:num_sign_indef}

We now turn to the case of sign-indefinite $\delta\alpha({\bf r})$.
In this section, we will study the complex eigenvalues of the matrix
$W_c$ defined in Section~\ref{subsec:sign_indef}. We note that, unlike
in the case of $W$ which is independent of the sign of $\delta\alpha$,
$W_c[-\delta\alpha] = - W_c[\delta\alpha]$. Note that the eigenvalues
of $W_c$ change sign when the sign of $\delta\alpha$ is inverted.

\begin{figure}
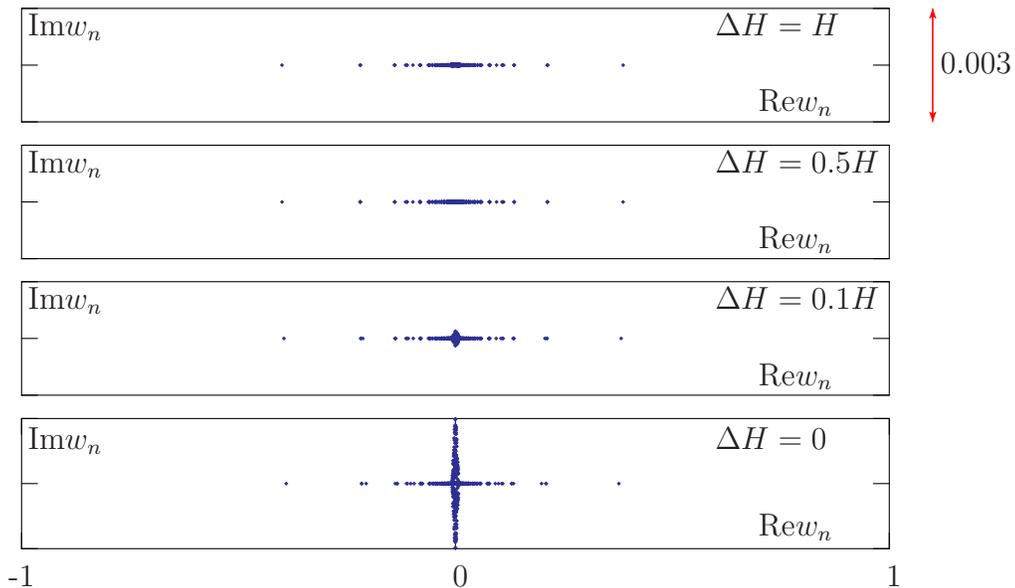

  \centerline{\input{fig5a.tex}}
\vspace*{-0.5cm}
  \centerline{\input{fig5b.tex}}
\vspace*{-0.5cm}
  \centerline{\input{fig5c.tex}}
\vspace*{-0.5cm}
  \centerline{\input{fig5d.tex}}
\caption{\label{fig:two_diff_cubes} All eigenvalues of the matrix $W_c$
  for two cubic inhomogeneities of equal sides $H=0.5\lambda_d$,
  placed side-by-side and separated by the surface-to-surface distance
  $\Delta H$. One cube has contrast $\kappa = +1$ and the other
  $\kappa=-1$. Discretization: $h=\lambda_d/20$.}
\end{figure}

The first example considered here is two cubic inhomogeneities similar
to those used to compute the data points for Fig.~\ref{fig:two_cubes},
but now one of them has the negative contrast $\kappa=-1$. In
Fig.~\ref{fig:two_diff_cubes}, all eigenvalues of $W_c$ for this
system are shown as dots in the complex plane.  When the cubes are
sufficiently far apart, the imaginary parts of the eigenvalues are
very small ($\sim 10^{-7}$ for $\Delta H = H$). This corresponds to
the non-interacting limit, when the interaction operator $W_c$ is,
approximately, block-diagonal, where each block is real symmetric. As
the cubes approach, some of the eigenvalues acquire imaginary parts.
The eigenstates with complex eigenvalues are ``hybridized'', i.e.,
they are collective eigenstates of the two interacting objects rather
than ``pure'' eigenstates of each object taken separately. However,
the hybridization is weak. Imaginary parts of the eigenvalues do not
exceed $0.0015$ in absolute value. Again, this is in agreement with
the idea of exponentially-suppressed long-range interactions.

\begin{figure}
  \centerline{\input{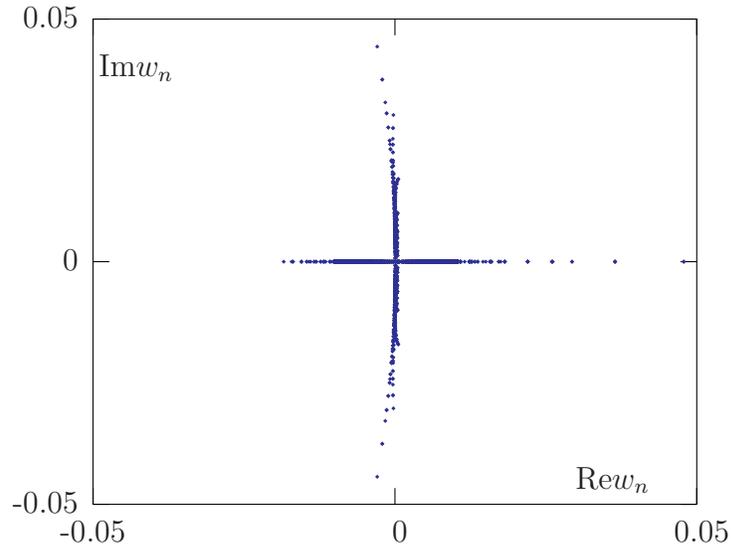}}
\caption{\label{fig:sandwich} All eigenvalues of the matrix $W_c$
  for the layered absorptive inhomogeneity described in the text.}
\end{figure}

Next, we consider a layered structure of fifteen thin square layers of
thickness $h$ and alternating contrast $\kappa = \pm 1$ sandwiched on
top of each other to form a cube of side $H=0.75\lambda_d$. The
discretization step is still $h=\lambda/20$. The eigenvalues of $W_c$
are shown in Fig.~\ref{fig:sandwich}. The displayed data indicate that
there are hybridized eigenstates (those with complex eigenvalues) and
eigenstates associated with an isolated thin layer and almost
unaffected by the interaction (with almost purely real eigenstates).
Overall, the absolute values of all eigenstates do not exceed $0.05$.
In this case, the matrix $W$ is negligibly small compared to $I$ and
can be neglected. This corresponds to the first Born approximation,
i.e., $T=V$. Thus, multiple scattering of diffuse waves for this
layered structure is quite weak and can be neglected with little loss
of precision.

\begin{figure}
  \centerline{\input{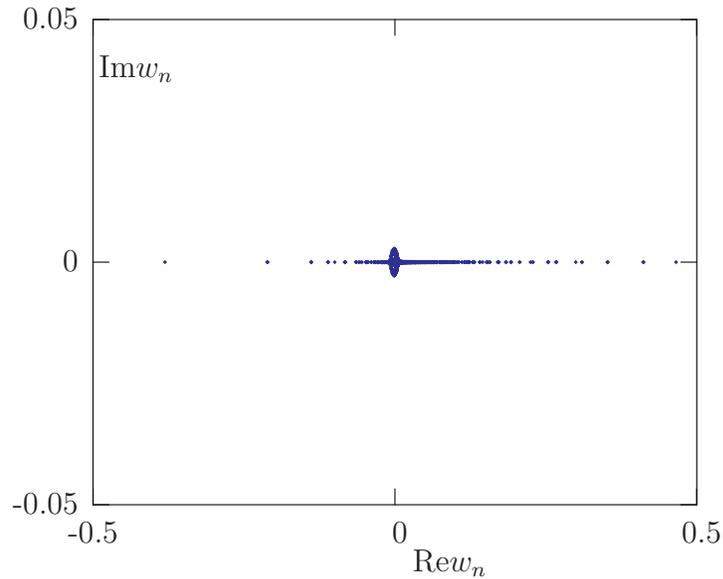}}
\caption{\label{fig:embedded_cubes} All eigenvalues of the matrix $W_c$
  for the absorptive inhomogeneity in the shape of two embedded cubes
  described in the text.} 
\end{figure}

The final example is one cubic inhomogeneity embedded inside another.
Namely, a cube of size $11h \times 11h \times 11h$ with
contrast $\kappa=-1$ was ``coated'' by a larger cube of size $21h
\times 21h \times 21h$ with contrast $\kappa=+1$. The contrasts in
the inner and outer cubes were not additive, so that $\kappa=-1$ in
the interior and $\kappa=+1$ in the exterior of the structure. The
discretization step was $h=\lambda_d/20$, so that the outer cube side
was $H_{\rm out} = 1.05\lambda_d$; the inner cube side was $H_{\rm in}
= 0.55\lambda_d$. The eigenvalues of the matrix $W_c$ for this structure
are shown in Fig.~\ref{fig:embedded_cubes}. Note that the vertical
scale in this figure is the same as in Fig.~\ref{fig:sandwich}, but the
horizontal scale is ten times larger. Thus, while multiple scattering
of diffuse waves inside each component (e.g., within the regions of
positive or negative contrast) is much stronger than in the case of
the layered structure of Fig.~\ref{fig:sandwich}, hybridization is
much weaker. The hybridized eigenvalues can be seen near the origin of
the complex plane and are all very small in magnitude. At the same
time, the eigenvalues that are relatively large in magnitude are
almost purely real, which is characteristic for weak interaction
between regions with positive and negative contrasts.

\section{Discussion}
\label{sec:discussion}

In this paper, we have derived a sufficient condition for convergence
of the Born series for the forward operator of optical tomography. The
condition is quite simple and states that the series converge if the
relative deviation of the absorption coefficient from its background
value $\delta\alpha({\bf r})/\alpha_0$ does not exceed unity,
independently of the support of $\delta\alpha({\bf r})$.  A similar
condition was obtained for scattering inhomogeneities which are
manifested by a spatially inhomogeneous diffusion coefficient. We have
considered absorbing and scattering inhomogeneities separately; the
situation when the absorption and the diffusion coefficients can vary
in space simultaneously is not discussed in this paper.  We argue that
the convergence condition depends only on the amplitude but not on
shape of $\delta\alpha$ (or $\delta D$) due to the exponential spatial
decay of diffuse waves. Because of this decay, multiple scattering is
suppressed on large scales. We emphasize again that we discuss here
multiple scattering of diffuse waves -- scalar solutions to the
diffusion equation (\ref{DE_1}) -- not electromagnetic multiple
scattering which happens at much smaller physical scales.  In the case
when $\delta\alpha({\bf r})$ has a compact support in a ball of radius
$a$, a sharper convergence condition has been obtained (formula
(\ref{2})), which is a generalization of the result previously
obtained for the scalar wave equation~\cite{colton_book_98}. A crucial
difference between the convergence condition for propagating and
diffuse waves is revealed in the limit $a\rightarrow \infty$, as is
discussed in Section~\ref{sec:colton}.

An interesting consequence of the convergence condition is that the
nonlinearity of the inverse problem of optical tomography can be
controlled if the constant $\alpha_0$ can be controlled. Thus,
increasing $\alpha_0$ results in effective linearization of the
inverse problem. Theoretically, $\alpha_0$ can be chosen arbitrarily.
However, the ill-posedness of the {\em linear} inverse problem tends
to increase with $\alpha_0$. This reveals an interplay between the
ill-posedness of the linearized inverse problem and the degree of
nonlinearity of the full inverse problem (before linearization). Note
that in experiments, $\alpha_0$ can be tuned, for example, by changing
the composition of an index-matching fluid.

We have performed numerical simulations for absorbing inhomogeneities.
All numerical data are in agreement with the analytical results of
this paper. We have found that the derived convergence condition is
satisfied for a very broad range of parameters which are accessible in
numerical experiments. We have also found that the effects of multiple
scattering between spatially separated inhomogeneities such as two
separate cubes is quite weak.  This is again a consequence of the
exponential decay of diffuse waves.  Interaction of inhomogeneities
whose contrasts have different signs was found to be especially weak.
Thus, for the layered structure discussed in
Section~\ref{subsec:num_sign_indef}, the interaction is insignificant
and the first Born approximation can be used with high accuracy --
even though the object is a layered cube of size $H=0.75 \lambda_d$.

While we have found no substantial interaction between spatially
separated inhomogeneities, nonlinearity can become strong in bulk
inhomogeneities of large spatial extent or high contrast. In this
case, the nonlinearity results from short-range interactions. Here two
voxels can strongly interact with each other even if they are far
apart, provided that there is a continuous path of other voxels
connecting them.

Another aspect of the paper that deserves comment is the independence
of the results on source-detector orientation.  Indeed, it may seem
natural that two absorbing cubes that block the line of sight will
have more effect on the measured signal than the same two cubes
rotated so that only one of them blocks the line of sight. In fact,
convergence or divergence of the Born series can be influenced, to a
certain extent, by the source-detector arrangement.  Indeed,
calculation of the measurable signal according to (\ref{Dyson_matrix})
involves multiplication of the T-matrix by $G_0^{\rm DV}$ and
$G_0^{\rm VS}$ from left and right. These matrices are source- and
detector-dependent. It can happen that the matrix $W$ has an
eigenvalue larger than unity so that the Born series for the T-matrix
diverges, but the corresponding eigenvector has a zero projection on
either $G_0^{\rm DV}$ or $G_0^{\rm VS}$. Then the Born expansion of
the Green's function $G^{\rm DS}$ will converge for the selected
source-detector configuration.  However, if the Born series converges
for the T-matrix, it converges for {\em all possible} source-detector
pairs.

Finally, our results pertain only to convergence of the {\em forward
  series}. Analogous results on the convergence of the inverse series
are not yet known.

\section*{Acknowledgment}

This research was supported by the NSF grants DMS-0554100 and
EEC-0615857.

\section*{References}
\bibliography{abbrev,master,book}

\end{document}